\def\~{{$\tilde{\phantom{a}}$}}

\documentclass [12pt] {article}
\usepackage{epsfig}
\usepackage{color}

\def\thebibliography#1{\section{References}\markboth
 {REFERENCES}{REFERENCES}\list
 {[\arabic{enumi}]}{\settowidth\labelwidth{[#1]}\leftmargin\labelwidth
 \advance\leftmargin\labelsep
 \usecounter{enumi}}
 \def\newblock{\hskip .11em plus .33em minus -.07em}
 \sloppy
 \sfcode`\.=1000\relax}
\def\upcite#1{\raise6pt\hbox{\scriptsize
\cite{#1}}}
  \def\lsim{\mathrel {\vcenter {\baselineskip 0pt \kern 0pt
    \hbox{$<$} \kern 0pt \hbox{$\sim$} }}}
    \def\gsim{\mathrel {\vcenter {\baselineskip 0pt \kern 0pt
    \hbox{$>$} \kern 0pt \hbox{$\sim$} }}}

\def\hline{\noalign{\hrule \vskip2pt}}

\def\|{\ifmmode\Vert\else \char`\|\fi}
\ifx\oldzeta\undefined                          
  \let\oldzeta=\zeta                            
  \def\zzeta{{\raise 2pt\hbox{$\oldzeta$}}}     
  \let\zeta=\zzeta                              
\fi

\ifx\oldchi\undefined                           
  \let\oldchi=\chi                              
  \def\cchi{{\raise 2pt\hbox{$\oldchi$}}}       
  \let\chi=\cchi                                
\fi

\def\frac#1#2{{#1 \over #2}}

\def\half{\ifinner {\scriptstyle {1 \over 2}}
   \else {1 \over 2} \fi}

\def\simge{\mathrel{%
   \rlap{\raise 0.511ex \hbox{$>$}}{\lower 0.511ex \hbox{$\sim$}}}}
\def\simle{\mathrel{
   \rlap{\raise 0.511ex \hbox{$<$}}{\lower 0.511ex \hbox{$\sim$}}}}

\def\buildchar#1#2#3{{\null\!                   
   \mathop#1\limits^{#2}_{#3}                   
   \!\null}}                                    
\def\overcirc#1{\buildchar{#1}{\circ}{}}

\def\slashchar#1{\setbox0=\hbox{$#1$}           
   \dimen0=\wd0                                 
   \setbox1=\hbox{/} \dimen1=\wd1               
   \ifdim\dimen0>\dimen1                        
      \rlap{\hbox to \dimen0{\hfil/\hfil}}      
      #1                                        
   \else                                        
      \rlap{\hbox to \dimen1{\hfil$#1$\hfil}}   
      /                                         
   \fi}                                         %

\def\subrightarrow#1{
  \setbox0=\hbox{
    $\displaystyle\mathop{}
    \limits_{#1}$}
  \dimen0=\wd0
  \advance \dimen0 by .5em
  \mathrel{
    \mathop{\hbox to \dimen0{\rightarrowfill}}
       \limits_{#1}}}                           

\def\overlay#1#2{\ifmmode%
\setbox0=\hbox{$#1$}%
\setbox1=\hbox to\wd0{\hss$#2$\hss}\else%
\setbox0=\hbox{#1}%
\setbox1=\hbox to\wd0{\hss#2\hss}\fi%
#1\hskip-\wd0\box1 }

\def\pmb#1{\leavevmode\setbox0=\hbox{#1}%
\kern-.02em\copy0\kern-\wd0
\kern.04em\copy0\kern-\wd0
\kern-.02em\raise.04em\box0 }

\def\vereq#1#2{\lower3pt\vbox{\baselineskip1.5pt \lineskip1.5pt
\ialign{$\m@th#1\hfill##\hfil$\crcr#2\crcr\sim\crcr}}}

\def\tensor#1{\protect\@ontopof{#1}{\leftrightarrow}{1.15}\mathord{\box2}}
\def\overstar#1{\protect\@ontopof{#1}{\ast}{1.15}\mathord{\box2}}
\def\overdots#1{\protect\@ontopof{#1}{\cdots}{1.0}\mathord{\box2}}
\def\overcirc#1{\protect\@ontopof{#1}{\circ}{1.2}\mathord{\box2}}
\def\loarrow#1{\protect\@ontopof{#1}{\leftarrow}{1.15}\mathord{\box2}}
\def\roarrow#1{\protect\@ontopof{#1}{\rightarrow}{1.15}\mathord{\box2}}

\def\@ontopof#1#2#3{%
{\mathchoice
{\@@ontopof{#1}{#2}{#3}\displaystyle\scriptstyle}%
{\@@ontopof{#1}{#2}{#3}\textstyle\scriptstyle}%
{\@@ontopof{#1}{#2}{#3}\scriptstyle\scriptscriptstyle}%
{\@@ontopof{#1}{#2}{#3}\scriptscriptstyle\scriptscriptstyle}%
}%
}

\def\@@ontopof#1#2#3#4#5{%
\setbox0=\hbox{$#4#1$}%
\setbox1=\hbox{$#5#2$}%
\setbox2=\hbox{}\ht2=\ht0 \dp2=\dp0 %
\ifdim\wd0>\wd1 %
\setbox1=\hbox to\wd0{\hss\box1\hss}%
\mathord{\rlap{\raise#3\ht0\box1}\box0}%
\else   %
\setbox1=\hbox to.9\wd1{\hss\box1\hss}%
\setbox0=\hbox to\wd1{\hss$#4\relax#1$\hss}%
\mathord{\rlap{\copy0}\raise#3\ht0\box1}%
\fi
}%

\def\lambdabar{\protect\@lambdabar}
\def\@lambdabar{%
\relax
\bgroup
\def\@tempa{\hbox{\raise.73\ht0
\hbox to0pt{\kern.25\wd0\vrule width.5\wd0
height.1pt depth.1pt\hss}\box0}}%
\mathchoice{\setbox0\hbox{$\displaystyle\lambda$}\@tempa}%
{\setbox0\hbox{$\textstyle\lambda$}\@tempa}%
{\setbox0\hbox{$\scriptstyle\lambda$}\@tempa}%
{\setbox0\hbox{$\scriptscriptstyle\lambda$}\@tempa}%
\egroup
}

\def\corresponds{{\lower.2ex\hbox{=}}{\rm\kern-.75em^\triangle}}
\def\succsim{\succ\kern-.9em_\sim\kern.3em}
\def\precsim{\prec\kern-1em_\sim\kern.3em}
\def\slantfrac#1#2{\kern1em^{#1}\kern-.3em/\kern-.1em_{#2}}

\def\twiddle{{$\tilde{\phantom{a}}$}}

\begin{document}

\begin{center}
{\Large\bf Noncontact measurement of the tension of a wire}
\\

\medskip

Mark R.~Convery\footnote{Present address: Stanford Linear Accelerator Center,
 Stanford, California 94309.}
and
Kirk T.~McDonald
\\
{\sl Joseph Henry Laboratories, Princeton University, Princeton, NJ 08544}
\\
(April 14, 1996)
\end{center}

\section{The Problem}

A conducting wire of mass $m$ and length $L$ is stretched between two fixed 
points with an unknown tension $T$.  One could determine the tension by 
plucking the wire
and observing the frequency of the vibration.  Analyze the following scheme
for noncontact plucking.  A magnetic field of strength $B$ is applied
transversely to the wire over a length $l \ll L$ around the midpoint of the
wire.  A very short pulse of total charge $Q$ is passes down the wire, which
therefore starts vibrating.  The voltage induced between the two ends of
the vibrating wire is measured as a function of time and a Fourier analysis
yields the various frequencies present.

\begin{figure}[htp]  
\begin{center}
\includegraphics[width=4in, angle=0, clip]{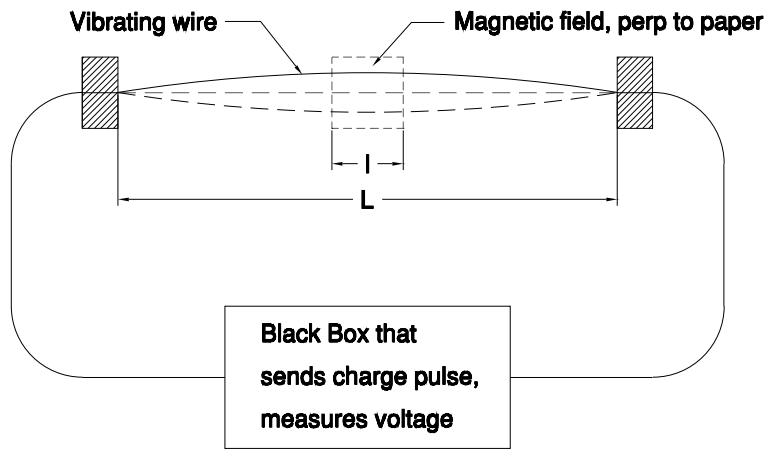}
\parbox{5.5in} 
{\caption[ Short caption for table of contents ]
{\label{tense} Scheme from noncontact measurement of wire tension.
}}
\end{center}
\end{figure}

Calculate the voltage induced at the various frequencies, and relate those
frequencies to the tension in the wire.  You may ignore damping effects.

A device constructed on the principles of this problem is described in
\cite{Convery}.

\section{Solution}

According to Faraday, the induced voltage is $V = -\dot \Phi/c$ in 
Gaussian units, where  $\Phi$ is the magnetic flux and $c$ is the speed of light.

We write the amplitude of the transverse oscillation of the wire as $a(x,t)$.
The magnetic flux through the circuit containing the wire is 
$\Phi \approx \Phi_0 + Bla(0,t)$, where $\Phi_0$ is the flux when the wire is at
rest.
taking the origin at the midpoint of the wire.  The boundary conditions on the
wire are $a(L/2,t) = 0 = a(-L/2,t)$, and one of the initial conditions is
$a(x,0) = 0$
at the moment the current pulse is applied.  These are sufficient to
determine the amplitude as having the form
\begin{equation}
a(x,t) = \sum_{{\rm odd}\ n} a_n \cos{n\pi x \over 2L} \sin\omega_n t.
\label{eq1}
\end{equation}
The motion satisfies the wave equation $a'' = \ddot a/v^2$,
where the wave velocity is related to the tension by
$v = \sqrt{TL/m}$.
Hence,
\begin{equation}
\omega_n = {n\pi v \over 2L} = {n\pi \over 2}\sqrt{T \over mL}.
\label{eq2}
\end{equation}
To complete the solution we need another initial condition, corresponding to
the transverse impulse due the current pulse interacting with the magnetic
field.  The Lorentz force on the length $l$ of wire in the magnetic field
due to current $I$ is $F = IBl/c$.
If this lasts for time $\Delta t$ the resulting impulse is
$\Delta P = I\Delta t Bl/c = QBl/c$.
Only a length $l$ of the wire experiences this impulse, so the mass of this
section is $ml/L$ and the initial velocity is $QBL/mc$.  In sum, the
second initial condition is
\begin{equation}
\dot a(x,0) = \left\{  \begin{array}{ll}
                         QBL/mc, & |x| < l/2; \\
                         0,      & |x| > l/2. 
\end{array} \right.
\label{eq3}
\end{equation} 

From the form (\ref{eq1}) of $a(x,t)$ we deduce that
\begin{equation}
\dot a(x,0) = \sum_{{\rm odd}\ n} a_n\omega_n \cos{n\pi x \over 2L}.
\label{eq4}
\end{equation}

Hence, in the usual manner we evaluate the Fourier coefficients as
\begin{equation}
a_n\omega_n = {2 \over L} \int_{-L/2}^{L/2} \dot a(x,0) \cos{n\pi x \over 2L} dx\approx {2QBl \over mc},
\label{eq5}
\end{equation}
for $l \ll L$.

The induced voltage is then
\begin{equation}
V(t) = -{Bl \over c} \dot a(0,t) = - {2 Q(Bl)^2 \over mc^2}
\sum_{{\rm odd}\ n} \cos\omega_n t.
\label{eq6}
\end{equation}
The amplitude of the voltage induced at frequency $\omega_n$ is
$2Q(Bl)^2/mc^2$, independent of $n$.
In practice, the finite values of $l$ and $\Delta t$ reduce the amplitudes of
the higher harmonics.

\end{document}